\documentclass[final,3p,times,12pt,twocolumn]{elsarticle}

\usepackage{graphicx}
\usepackage{amssymb}

\journal{Icarus}

\begin{document}

\begin{frontmatter}


\title{The Transient Jupiter Trojan-Like Orbit of P/2019 LD$_2$ (ATLAS)}



\author[psi,asiaa]{Henry H.\ Hsieh}
\author[qub]{Alan Fitzsimmons}
\author[ubel]{Bojan Novakovi\'c}
\author[uh]{Larry Denneau}
\author[uh]{Aren N.\ Heinze}

\address[psi]{Planetary Science Institute, 1700 East Fort Lowell Rd., Suite 106, Tucson, AZ 85719, USA; hhsieh@psi.edu}
\address[asiaa]{Institute of Astronomy and Astrophysics, Academia Sinica, P.O.\ Box 23-141, Taipei 10617, Taiwan}
\address[qub]{Astrophysics Research Centre, School of Mathematics and Physics, Queen’s University Belfast, Belfast BT7 1NN, UK}
\address[ubel]{Department of Astronomy, Faculty of Mathematics, University of Belgrade, Studentski trg 16, 11000 Belgrade, Serbia}
\address[uh]{Institute for Astronomy, University of Hawaii, 2680 Woodlawn Drive, Honolulu, Hawaii 96822, USA}

\begin{abstract}
Comet P/2019 LD$_2$ has orbital elements currently resembling those of a Jupiter Trojan, and therefore superficially appears to represent a unique opportunity to study the volatile content and active behavior of a member of this population for the first time. However, numerical integrations show that it was previously a Centaur before reaching its current Jupiter Trojan-like orbit in 2018 July, and is expected to return to being a Centaur in 2028 February, before eventually becoming a Jupiter-family comet in 2063 February.  The case of P/2019 LD$_2$ highlights the need for mechanisms to quickly and reliably dynamically classify small solar system bodies discovered in current and upcoming wide-field surveys.
\looseness=-1
\end{abstract}


\end{frontmatter}


\section{Background} \label{sec:intro}

Jupiter Trojans are small solar system bodies that share Jupiter's orbit around the Sun and reside in one of two ``clouds'' associated with the stable L4 and L5 Lagrange regions located 60$^{\circ}$ ahead of and behind the planet in its orbit \citep[see][]{emery2015_trojans_ast4a_short}.  Their origins are currently uncertain, with potential scenarios under debate including formation near their current locations and capture by Jupiter from source regions farther out in the solar system \citep[e.g.,][]{marzari2002_trojanorigins_asteroids3a,morbidelli2005_trojancapture_short,slyusarev2014_jupitertrojans,pirani2019_planetarymigrationsmallbodies_short}.  
Observational studies have shown the population to consist primarily of C-, P-, and D-type asteroids \citep[cf.][]{grav2012_neowisetrojans}, where measurements of low densities for some objects indicate that they could be highly porous, volatile-rich, or both \citep[e.g.,][]{marchis2006_patroclusdensity_short,marchis2014_hektor_short}.
Meanwhile, thermal models have shown that H$_2$O ice could remain preserved on Jupiter Trojans over the age of the solar system under just $\sim$10~cm of dust at their poles to $\sim$10~m of regolith elsewhere \citep{guilbert-lepoutre2014_trojanice}.  Thus, cometary activity could be possible on Trojans, perhaps triggered by impacts and driven by hypervolatile species like CO or CO$_2$ \citep[cf.][]{womack2017_cometsCO}. No active Trojans have been reported to date, however.
\looseness=-1

\begin{figure}[tbh]
\centerline{\includegraphics[width=3.2in]{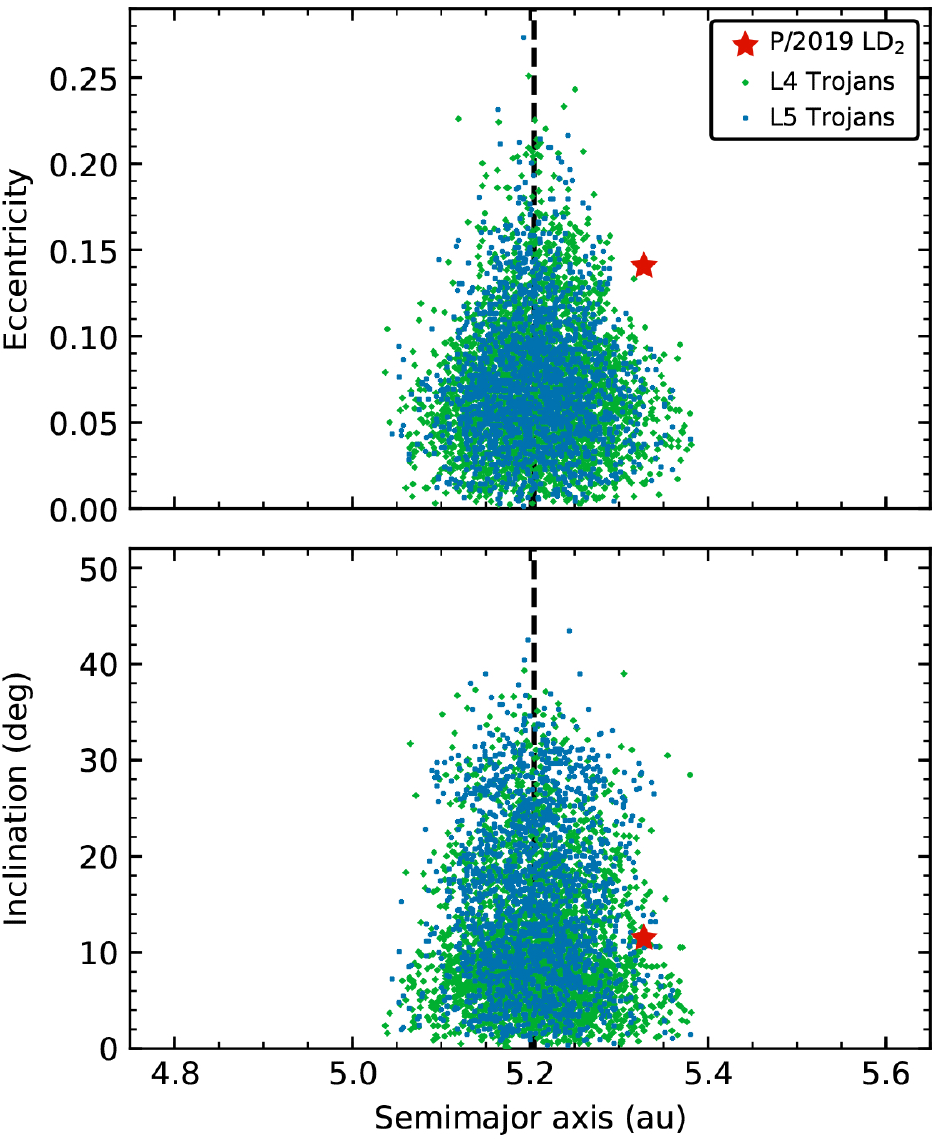}}
\caption{\small Osculating semimajor axis versus osculating eccentricity (top) and osculating inclination (bottom) plots of all numbered Jupiter Trojans (as of 2020 May 25), where objects in the L4 and L5 Trojans clouds are marked by small green and blue dots, respectively, and P/2019 LD$_2$ is marked by a red star.  
}
\label{figure:aei_plot}
\end{figure}

P/2019 LD$_2$ was discovered
on UT 2019 June 10 at a heliocentric distance of $R=4.666$~au by the 0.5-m Asteroid Terrestrial-Impact Last Alert System telescope on Mauna Loa in Hawaii. 
Suspected cometary activity in discovery images analyzed by the ATLAS team was confirmed by followup observations on 2019 June 11, 13, and 29
\citep{fitzsimmons2020_p2019ld2,sato2020_p2019ld2_short}.
The object currently has Jupiter Trojan-like orbital elements (Figure~\ref{figure:aei_plot}), with a semimajor axis of $a=5.3279$~au, eccentricity of $e=0.1407$, and inclination of $i=11.517^{\circ}$.  If P/2019 LD$_2$ is in fact a Jupiter Trojan, it would represent a unique opportunity to study the volatile content and behavior of a member of this population of objects for the first time and to use the results of those investigations to constrain models of solar system formation and evolution.
\looseness=-1

\begin{figure}[tbh]
\centerline{\includegraphics[width=3.2in]{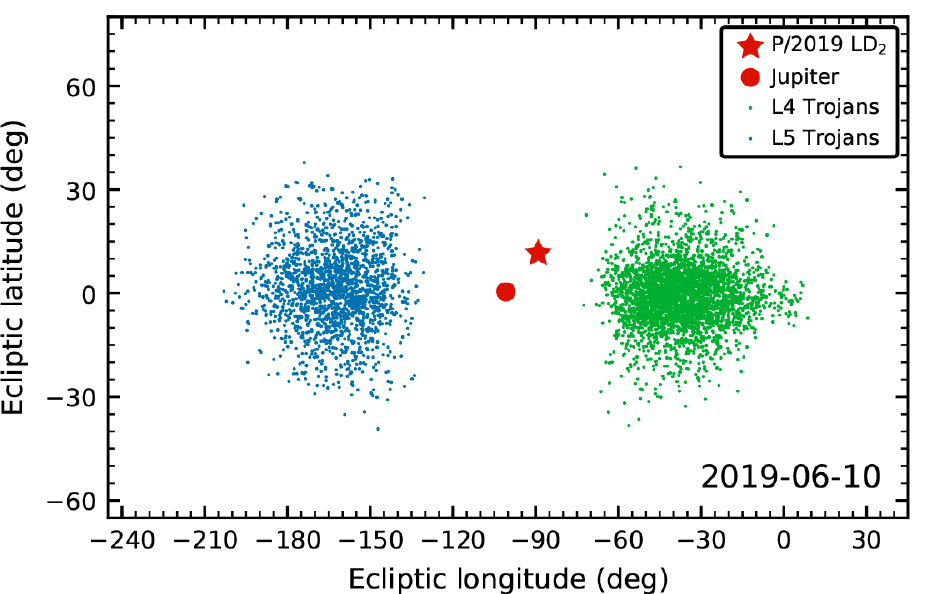}}
\caption{\small Plot of the positions of all numbered objects in Jupiter's L4 Trojan cloud (green dots) and L5 cloud (blue dots), Jupiter (red circle), and P/2019 LD$_2$ (red star) in heliocentric ecliptic latitude and longitude space on UT 2019 June 10.
\looseness=-1}
\label{figure:eclip_latlong_plot}
\end{figure}

However, a heliocentric ecliptic latitude and longitude plot of P/2019 LD$_2$ and other Jupiter Trojans at the time of the object's discovery (Figure~\ref{figure:eclip_latlong_plot}) gives an indication that P/2019 LD$_2$ might not be a true Jupiter Trojan, as it much closer to Jupiter in ecliptic longitude ($\sim$10$^{\circ}$) than any other Jupiter Trojans ($\sim$40$^{\circ}$-$100^{\circ}$) and does not clearly belong to either the L4 or L5 clouds.
Dynamical analyses reported in \citet{sato2020_p2019ld2_short} suggest that P/2019 LD$_2$'s orbital elements are unstable, inconsistent with the behavior expected of a true Jupiter Trojan, while similar analyses disputing P/2019 LD$_2$'s classification as a Jupiter Trojan were reported by Sam Deen and Tony Dunn in posts to the Minor Planet Mailing List\footnote{{\tt https://groups.io/g/mpml}} (MPML; an email discussion list for the amateur and professional minor planet community) as well as by \citet{kareta2020_2019ld2_short}.  In this note, we present numerical integration results confirming and characterizing the non-Trojan-like dynamical behavior of P/2019 LD$_2$ and briefly discuss the implications of this object for current and future surveys.
\looseness=-1

%
%

\section{Experimental Design} \label{sec:expdesign}

To assess P/2019 LD$_2$'s dynamical nature, we generated 100 dynamical clones drawn from the multivariate normal distribution for the object (as of UT 2020 June 1), defined by an orbital covariance matrix, provided by the JPL Small Bodies Database\footnote{\tt https://ssd.jpl.nasa.gov/sbdb.cgi}.  Dynamical clones are used here to assess the amount of potential divergence due to chaos in P/2019 LD$_2$'s predicted orbital evolution that could occur due to the object's orbital element uncertainties.
We also performed the same procedure for six reference Jupiter Trojans: (588) Achilles, (624) Hektor, and (659) Nestor from Jupiter's L4 Trojan cloud and (617) Patroclus, (884) Priamus, and (1172) Aneas from the L5 cloud.  We then conducted backward and forward numerical integrations for all objects and their clones for 1000 years in each direction, using the Bulirsch-St{\"o}er integrator in the Mercury $N$-body integration package \citep{chambers1999_mercury}.  To study the long-term stability of P/2019 LD$_2$, we also conducted forward integrations for 1 Myr for all test particles.  All integrations accounted for gravitational perturbations from the seven major planets except for Mercury and used an initial time step of 0.1 days.  In all integrations, particles are removed when they reach $a$$\,>\,$100~au.  Non-gravitational forces were not included.
\looseness=-1

\section{Results and Analysis} \label{sec:Analysis}

\begin{figure*}[tpbh]
\centerline{\includegraphics[width=6.5in]{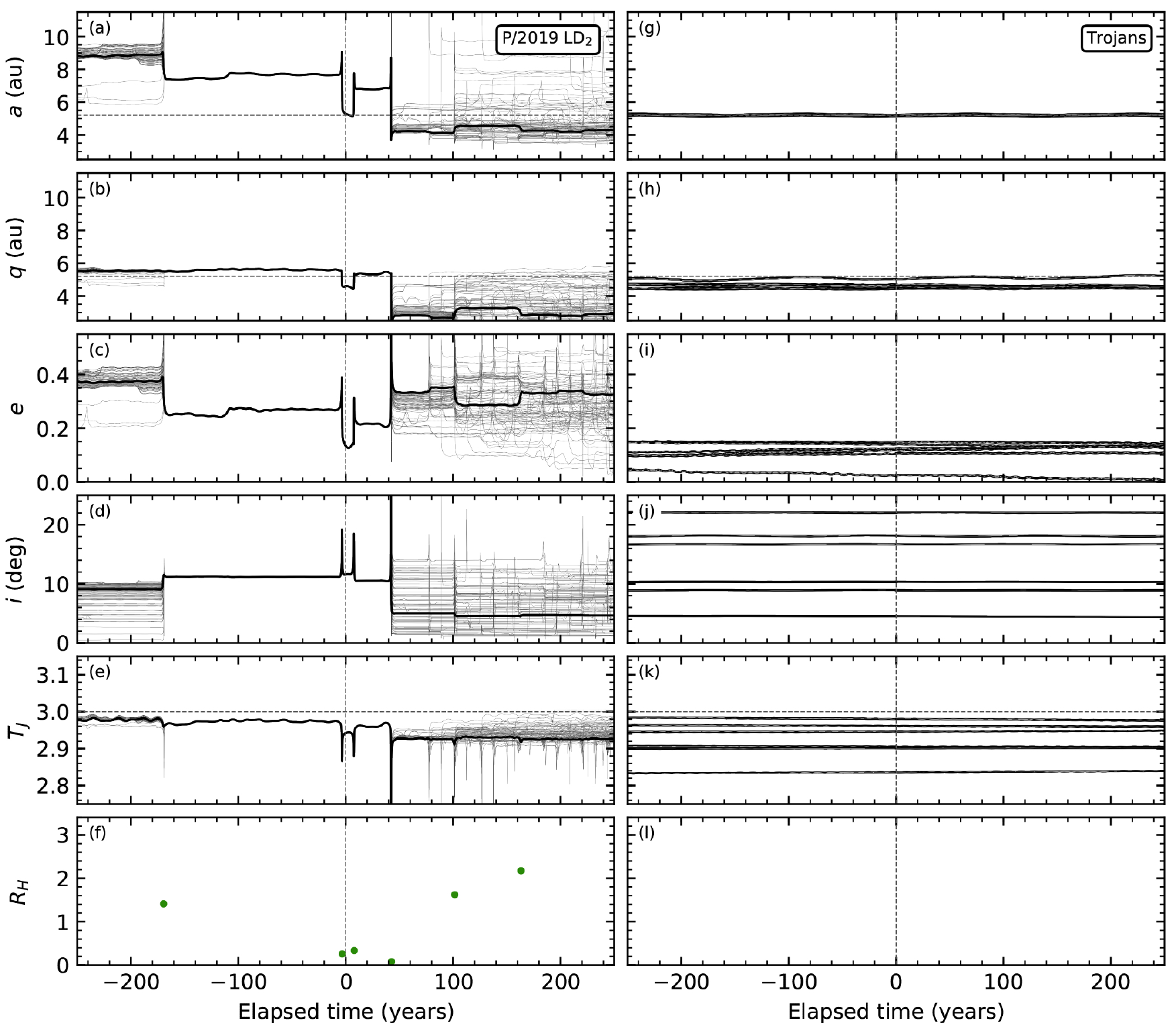}}
\caption{\small Plots of the backward (negative elapsed time) and forward (positive elapsed time) dynamical evolution of P/2019 LD$_2$ (left panels) and six reference Jupiter Trojans (right panels).  Orbital parameters plotted as functions of time, from top to bottom, are semimajor axis, perihelion distance, eccentricity, inclination, Tisserand parameter ($T_J$), and distances of close encounters with Jupiter (green dots; shown for P/2019 LD$_2$ only) in terms of the planet's Hill radius ($R_{H,{\rm Jup}}$). In panels (a) to (e), thick and thin black lines mark the evolution of P/2019 LD$_2$ itself and its dynamical clones, respectively. In panels (g) to (l), thick lines mark the evolution of each reference Trojan, where the evolution of all dynamical clones is effectively identical to their corresponding real objects.  Vertical dashed lines in all panels mark the present day.  Horizontal dashed lines in panels (a), (b), (g), and (h) mark the semimajor axis of Jupiter, while horizontal dashed lines at $T_J=3$ in panels (e) and (k) mark the canonical dividing line between dynamically asteroidal and dynamically cometary orbits.
}
\label{figure:orbelem_evolution}
\end{figure*}

Results from our backward and forward 1000-year integrations are plotted in Figure~\ref{figure:orbelem_evolution}. We can confirm the assessments reported in \citet{sato2020_p2019ld2_short} and \citet{kareta2020_2019ld2_short} that P/2019 LD$_2$ is only temporarily in a Jupiter Trojan-like orbit, where we find its overall dynamical behavior to be that of an active Centaur \citep[e.g.,][]{jewitt2009_actvcentaurs} transitioning into a Jupiter-family comet (JFC).
We find that P/2019 LD$_2$'s semimajor axis became Jupiter Trojan-like ($5.0~{\rm au}< a < 5.4$~au) in 2018 July and will remain in that range until 2028 February. The transitions into and out of P/2019 LD$_2$'s current orbit correspond to close encounters with Jupiter for all P/2019 LD$_2$-associated test particles (i.e., the object itself as well as all of its dynamical clones) in our integrations when the object passed within 0.09~au (or $0.25R_{H,{\rm Jup}}$, where $R_{H,{\rm Jup}}=0.355$~au is Jupiter's Hill radius) from Jupiter on 2017 February 20, and will pass within 0.12~au ($0.34R_{H,{\rm Jup}}$) from Jupiter on 2028 May 12.
\looseness=-1



Immediately prior to reaching its current Jupiter Trojan-like orbit in 2018 July, P/2019 LD$_2$'s orbital elements (Figure~\ref{figure:orbelem_evolution}) placed it in a Centaur-like orbit.  Here, we use the definition from \citet{jewitt2009_actvcentaurs}, where an object is considered a Centaur if $a_J < q < a_N$, $a_J < a < a_N$, and it is not in a 1:1 mean-motion resonance with any planet, where $a_J=5.204$~au and $a_N=30.047$~au are the semimajor axes of Jupiter and Neptune, respectively.  P/2019 LD$_2$ is expected to return to a Centaur-like orbit in 2028 February and remain there until 2063 February, when a very close encounter with Jupiter at $\sim$0.03~au ($\sim$0.08$R_{H,{\rm Jup}}$) in 2063 January will lower both its semimajor axis and perihelion distance to well below $a_J$, at which point, the object will be considered a JFC.  For comparison, integrations of our reference Trojans indicate that they remain on stable orbits for the duration of both our backward and forward 1000-yr integrations.  While the eccentricities of some of these objects drift smoothly over time and $a$ and $q$ exhibit small oscillations (Figure~\ref{figure:orbelem_evolution}g,h,i), we see none of the sharp orbital element changes exhibited by P/2019 LD$_2$.
\looseness=-1

We find that the orbital evolution trajectories of P/2019 LD$_2$ and all of its dynamical clones in our integrations between 1851 and 2063 are nearly identical, suggesting that our results likely reliably capture P/2019 LD$_2$'s true orbital evolution during this period.  However, before and after this time period, which is bracketed by close encounters with Jupiter at distances of $\sim$0.5~au ($1.4R_{H,{\rm Jup}}$) in 1850 November and $\sim$0.03~au ($\sim$0.08$R_{H,{\rm Jup}}$) in 2063 January, trajectories from our integrations for P/2019 LD$_2$ and its dynamical clones diverge widely (Figure~\ref{figure:orbelem_evolution}).  This divergence is a result of the chaotic nature of P/2019 LD$_2$'s orbit, especially during close encounters with Jupiter, and indicates that predictions about the object's dynamical behavior before 1851 or after 2063 should be regarded as highly uncertain.  Consideration of non-gravitational perturbations due to cometary outgassing could introduce even more uncertainty to our analysis of P/2019 LD$_2$'s orbital evolution, but given the expected weakness of any cometary activity at these large heliocentric distances, we expect outgassing perturbations to be essentially negligible compared to effects from the close encounters with Jupiter discussed above.
\looseness=-1


In our 1~Myr forward integrations of its nominal orbit, P/2019 LD$_2$ reaches $a$$\,>\,$100~au (and is removed from the integrations) in $9.9$$\,\times\,$10$^5$~yr (with many significant orbital element changes during that time), while all but three of its dynamical clones also reach $a$$\,>\,$100~au within 1~Myr with a median lifetime of $1.1$$\,\times\,$10$^5$~yr. This dynamical evolutionary behavior is consistent with current short-period comets \citep{hsieh2016_tisserand}, and contrasts sharply with our six reference Trojans, all of which remain in effectively the same orbits for the full 1~Myr integrations, further highlighting the dynamical distinction between P/2019 LD$_2$ and true Jupiter Trojans.
\looseness=-1

\section{Discussion} \label{sec:Discussion}


Despite the findings described above, it is possible that P/2019 LD$_2$ could have been a true Jupiter Trojan in the past and was driven onto its current orbit by non-gravitational perturbations arising from its cometary activity or other effects.  Jupiter Trojans are expected to occasionally escape from their stable orbits due to chaotic diffusion or collisions, and potentially contribute to other populations such as Centaurs and JFCs \citep[e.g.,][]{horner2012_anchises,disisto2019_trojanevolution}, and cometary non-gravitational perturbations could certainly accomplish similar effects \citep{yeomans2004_cometnongravs_comets2a_short}.  Given the object's clearly un-Trojan-like recent orbital history, however, we consider this to be an implausible scenario. There is no reason that P/2019 LD$_2$'s current transient resemblance to Jupiter Trojans should suggest that it is necessarily more likely than any other Centaur to have been a Jupiter Trojan in the past.  
Nonetheless, a future analysis of escape trajectories from the Trojan clouds involving non-gravitational perturbations due to cometary outgassing could be useful for assessing the potential contribution of active Jupiter Trojans to the Centaur and JFC populations.
In the meantime, observational characterization of P/2019 LD$_2$'s surface to determine if it has the ultrared colors of other Centaurs \citep[][]{jewitt2018_trojancolors} or has C-, P-, or D-type colors similar to other Jupiter Trojans would be very useful for confirming the object's true origin.
\looseness=-1

While P/2019 LD$_2$ is not a true Jupiter Trojan, its discovery is nonetheless instructive. Current surveys like ATLAS will continue to discover active objects, some of which may belong to populations not previously known to exhibit activity, and upcoming surveys like the Vera C.\ Rubin Observatory's Legacy Survey of Space and Time (LSST) promise to discover even more \citep{schwamb2018_lsstsssc_short}.  The temporary capture of objects onto Trojan-like orbits is not expected to be frequent, but also not exceedingly rare \citep[e.g.,][]{karlsson2004_temporarytrojans,horner2006_centaurcapture}, where we note that temporary satellite captures could also be found to have nominally Trojan-like orbital elements \citep[e.g.,][]{tancredi1990_helinromancrockett,benner1995_shoemakerlevy9,ohtsuka2008_147p}.  As such, more cases like P/2019 LD$_2$ should be expected in the future.
\looseness=-1

Computationally scalable approaches for accurately dynamically classifying objects of interest discovered by wide-field surveys in a timely manner will be needed for both large-scale population studies and investigations of individual targets, especially as discovery rates increase \citep[cf.][]{hsieh2019_lsstsoftware_short,schwamb2019_lsstsoftware_short}.  To identify true Jupiter Trojans, possible approaches include using Lyapunov Characteristic Exponent (LCE) values or proper orbital elements to ensure that a given object is in a stable 1:1 mean-motion resonance with Jupiter \citep[e.g.,][]{melita2008_trojanstability_short,holt2020_trojanstability_short}.  Both proper orbital elements and LCE values are currently provided by the AstDyS-2 website\footnote{\tt https://newton.spacedys.com/astdys/} for all numbered and multi-opposition Jupiter Trojans, and preparations are being made to continue doing so in the LSST era \citep{hsieh2019_lsstsoftware_short}, although their computation typically requires relatively high-quality orbits. For newly discovered objects of high interest that have lower-quality orbits, it may be useful to develop mechanisms for performing more rapid preliminary dynamical analyses using $N$-body integrations as was done in this work (perhaps also allowing for cometary non-gravitational perturbations).
\looseness=-1

While recent sharp decreases in P/2019 LD$_2$'s semimajor axis and perihelion distance are perhaps the most plausible trigger of its current activity \citep{kareta2020_2019ld2_short,fernandez2018_centaurevolution}, tidal disruption or resurfacing \citep[e.g.,][]{binzel2010_neoearthencounters_short,hyodo2016_centaurrings} from the object's close encounter with Jupiter in 2017 could also have contributed to making cometary activity more likely by disrupting surface material and excavating buried surface ice.  In this regard, systematic $N$-body integration analyses could also be useful for identifying small bodies that have experienced recent close planetary encounters so that they can be monitored for possible cometary activity.

\section*{Acknowledgements} \label{sec:acknowledgements}

We thank K.\ Walsh for helpful discussion, and J.\ Horner and another anonymous reviewer for helpful comments that improved this work.  HHH acknowledges support from NASA via the Solar System Workings and Early Career Fellowship programs (Grants 80NSSC17K0723 and 80NSSC18K0193) as well as Solar System Exploration Research Virtual Institute (SSERVI) Cooperative Agreement grant NNH16ZDA001N. AF acknowledges support from UK STFC grant ST/P0003094/1.
LD and ANH acknowledge support from NASA via grants NN12AR55G,\linebreak 80NSSC18K0284, and 80NSSC18K1575 supporting the Asteroid Terrestrial-impact Last Alert System (ATLAS) project.
\looseness=-1







\bibliography{main.bib}
\bibliographystyle{icarus}\biboptions{authoryear}







\end{document}